\documentclass[amsmath,amssymb,aps,prl,superscriptaddress,twocolumn,floatfix]{revtex4}
\usepackage{graphicx}
\usepackage{dcolumn}

\usepackage[usenames]{color}
\usepackage{comment}
\usepackage{bm}
\usepackage{lipsum}
\usepackage{hyperref} 
\usepackage{physics}
\usepackage{bookmark}
\usepackage{mathtools}
\usepackage{subfig}

\begin{document}
\title{Unsupervised learning of two-component nematicity from STM data on magic angle bilayer graphene}
\author{William Taranto}
\affiliation{Laboratory of Atomic and Solid State Physics, Cornell University, 142 Sciences Drive, Ithaca NY 14853-2501, USA}
\author{Samuel Lederer}
\affiliation{Department of Physics, University of California, Berkeley, USA}

\author{Youngjoon Choi}

\affiliation{T. J. Watson Laboratory of Applied Physics, California Institute of Technology, Pasadena, CA, USA}
\affiliation{Institute for Quantum Information and Matter, California Institute of Technology, Pasadena, CA, USA}
\affiliation{Department of Physics, California Institute of Technology, Pasadena, CA, USA}

\author{Pavel Izmailov}
\affiliation{Courant Institute of Mathematical Sciences, New York University, NY, USA}
\author{Andrew Gordon Wilson}
\affiliation{Courant Institute of Mathematical Sciences, New York University, NY, USA}

\author{Stevan Nadj-Perge}
\affiliation{T. J. Watson Laboratory of Applied Physics, California Institute of Technology, Pasadena, CA, USA}
\affiliation{Institute for Quantum Information and Matter, California Institute of Technology, Pasadena, CA, USA}

\author{Eun-Ah Kim}
\affiliation{Laboratory of Atomic and Solid State Physics, Cornell University, 142 Sciences Drive, Ithaca NY 14853-2501, USA}

\begin{abstract}

Moir\'e materials such as magic angle twisted bilayer graphene (MATBG) exhibit remarkable phenomenology, but present significant challenges for certain experimental methods, particularly scanning probes such as scanning tunneling microscopy (STM). Typical STM studies that can image tens of thousands of atomic unit cells can image roughly ten moir\'e cells, making data analysis statistically fraught. Here, we propose a method to mitigate this problem by aggregating STM conductance data from several bias voltages, and then using the unsupervised machine learning method of gaussian mixture model clustering to draw maximal insight from the resulting dataset. We apply this method, using as input coarse-grained bond variables respecting the point group symmetry, to investigate nematic ordering tendencies in MATBG for both charge neutral and hole-doped samples. For the charge-neutral dataset, the clustering reveals the surprising coexistence of multiple types of nematicity that are unrelated by symmetry, and therefore generically nondegenerate. By contrast, the clustering in the hole doped data is consistent with long range order of a single type. Beyond its value in analyzing nematicity in MATBG, our method has the potential to enhance understanding of symmetry breaking and its spatial variation in a variety of moir\'e materials.
\end{abstract} 

\maketitle
  Magic angle twisted bilayer graphene (MATBG) and other moir\'e systems have emerged as highly tunable platforms for the experimental investigation of strongly correlated quantum matter\cite{Andrei2020NatMat,Andrei2021NatRevMat}. Rapid developments in multilayer graphitic systems have discovered rich phenomenology, including correlated insulating behavior \cite{Cao2018Nature_insulator}, topological phases\cite{Nuckolls2020Nature,Serlin2020Science,Choi2021,Saito2021,Das2021}, superconductivity\cite{Cao2018Nature,Park2021Nature}, and symmetry breaking such as orbital ferromagnetism\cite{Sharpe2019Science,Lu2019Nature} and other forms of flavor symmetry breaking\cite{Zondiner2020Nature,Wong2020Nature}, as well as nematicity\cite{Jiang2019Nature,Kerelsky2019Nature,Choi2019NatPhys,Cao2021Science,Rubio-Verdu2022,Samajdar20212DMaterials}. By virtue of its surface sensitivity and its spatial and energy resolution, scanning tunneling microscopy (STM) is one of the tools best suited to explore this physics. Indeed, STM studies have led to remarkable insights into local symmetry breaking tendencies in other materials\cite{Fischer2007RevModPhys}. However,
 STM studies of  moir\'e systems are challenging due to the large parameter space (the gate voltage, i.e. doping, being a new tuning parameter), as well as the large effective lattice constant, sketched in Fig. 1a . Because the lattice constant is so large, only a few unit cells are contained in a typical STM field of view (Figs. 1b,c), making edge effects significant. As a result, previous STM imaging studies on moir\'e systems have not explored spatial variations of symmetry breaking tendencies. 
 
  STM studies of moir\'e materials frequently observe the breaking of the $C_3$  rotational symmetry of the moir\'e lattice, as is strikingly illustrated in the STM conductance image of Fig. 1b. 
Rotational symmetry breaking, known as nematicity, is widely observed in correlated materials and increasingly provides a unifying thread among them \cite{Fradkin2010AnnRev,Fernandes2019AnnRev}. While the symmetry broken in the most commonly studied materials is a $C_4$ rotation, the $C_3$ rotational symmetry of  the moir\'e lattice introduces new aspects to the theoretical understanding of nematic order\cite{Xu2020PRB,Fernandese2020SciAdv,Samajdar20212DMaterials}. Specifically, the nematic order parameter for $C_3$ symmetry breaking transforms under a two-dimensional representation of the $C_3$ point group. Hence, two inequivalent types of nematic order can develop: type-I, with one strong direction, and type-II, with one weak direction, as shown in Fig. 1d (see supplementary material section II). While the existence of these two types is guaranteed by symmetry, little is known about the microscopic conditions that favor each type. Moreover, no experimental studies to date have explored this additional nematic degree of freedom in moir\'e materials. 

  Here, we propose an unsupervised machine learning approach to detect the two types of nematicity and their spatial fluctuations from STM data on moir\'e materials. Lately, the community has made rapid  progress in applying machine learning\cite{Carrasquilla2020AdvPhys0,Carleo2019RevModPhys} to experimental data on quantum matter from bulk probes such as resonant ultrasound\cite{Ghosh2020SciAdv}, neutron scattering\cite{Samarakoon2020NatComm}, and X-ray scattering\cite{Venderley2021} and microscopic probes such as STM\cite{Ziatdinov2016Nanotechnology,Zhang2019Nature,Cheung2020NatComm}, electron microscopy\cite{Cao2020Micro}, and quantum gas microscopy\cite{Rem2019NatPhys,Bohrdt2019NatPhys,Miles2021NatComm}. 
However, much of the literature has focused on supervised machine learning. One advantage of our technique is that, because it is unsupervised, it requires no training data, which is a significant advantage given the extraordinary difficulty of STM experiments. In previous work, this difficulty has been overcome using synthetic training data \cite{Zhang2019Nature}, but generating these data inevitably introduces bias. 
Unsupervised learning approaches look for naturally occurring clusters in the distribution of data\cite{Venderley2021} and enable unbiased discoveries. A key challenge in a successful implementation of an unsupervised learning method is defining the feature space for clustering. Since we are interested in phenomena at the moire lattice scale, we introduce coarse-grained features centered at each moire lattice cite (see Fig. 1f). We then study spatial variations in type-I and type-II nematic tendencies by clustering the features. 

The STM data used in this work have been obtained in the same 
experimental run as in Ref. \cite{Choi2019NatPhys}. MATBG is placed on an atomically smooth 
hexagonal 
boron nitride dielectric with the graphene gate underneath,
Fig. 1a. In this configuration, MATBG can be doped 
electrostatically i.e., by simply changing the gate voltage.
The spectroscopic maps used for investigating potential nematic 
phases were obtained at various doping levels and in close 
vicinity ($\pm20$~meV) of the Fermi level.

  Our unsupervised learning approach consists of two steps: feature selection and clustering. The goal of the feature selection step is to define a nematic order parameter centered at each moir\'e  lattice site. There are two complementary ways to describe nematic order in this setting, one borrowed from the physics of liquid crystals and the other from the theory of finite groups.
Nematic liquid crystals are described using a headless vector, the \emph{director} field $\eta=|\eta|(\cos\theta,\sin\theta)$, specifying the orientation of the long axis of the constituent molecules. For electronic liquid crystals on a lattice with $C_3$ symmetry, $\eta$ specifies an analogous elongation of local properties, and would naturally fall along high-symmetry lines--either along bond directions or mid-way between them. We dub these scenarios type I and type II nematicity respectively, as illustrated in Fig. 1d. From the group theory perspective, electronic nematicity can be described using a two-component order parameter $\bf \Phi$ transforming under the $E_2$ (or, informally, $[d_{x^2-y^2},d_{xy}]$) representation of $D_6$. These perspectives are equivalent, according to
\begin{equation}
{\bf\Phi}=|\eta|^2\begin{pmatrix}
\cos[2\theta]\\
\sin[2\theta]
\end{pmatrix}
\label{eq:polar}.
\end{equation}
We extract a local nematic order parameter from the STM data by defining three coarse-grained bond variables $\alpha,\beta,\gamma$ as shown in Fig 1f, and constructing the order parameter $\bf\Phi$ as

\begin{align}
\Phi_1=&\alpha-\frac{1}{2}\big[\beta+\gamma\big], \label{eq:phi1}\\
\Phi_2=&\frac{\sqrt{3}}{2}\big[\gamma-\beta\big]. \label{eq:phi2}
\end{align}
See supplementary material section 2 for further details. 

  Once we have collected the values of $\bf \Phi$ for each moir\'e lattice site and each bias voltage, we proceed to the clustering step. We combine order parameter values from selected voltage windows into a larger dataset for improved statistics, and then quantify the distribution of these data using gaussian mixture modeling (GMM)\cite{Bishop2006}. 
In GMM, each data point is assumed to have been drawn, with some probability (or ``weight") from one of several gaussians. The means and variances of the conjectured gaussian distributions, as well as their weights, are learned by maximizing the likelihood of the dataset being drawn from the model.  The clustering can quantify two distinct phenomena regarding local nematicity: spatial coexistence of different nematic tendencies, be it orientation or type, and long-range correlation of one nematic tendency. Separation between different clusters will signify multiple nematic tendencies coexisting within the field of view (see Fig. 1g-h). The spatial distribution of each cluster will establish correlation length of each nematic tendency. On the other hand, overlap among the clusters will signify long-range (limited by the field-of-view) correlation of one nematic tendency (see Fig. 1i-j).

 We first apply our analysis to data at charge neutrality point, with results shown in Fig. 2.
We focus on the vicinity of zero bias, specifically, the regions marked by gray strips in Fig. 2a, which shows the field-of-view averaged conductance as a function of bias voltage. In this region the sample exhibits the coexistence of different types of nematicity, as well as rapid variation of nematicity with bias voltage. At small positive voltages (rightmost strip of Fig. 2a) the clustering in the two-component order parameter space $[\Phi_1,\Phi_2]$ (Fig. 2b) yields results similar to the clustering of the synthetic data of Fig. 1g,h: the two clusters with appreciable populations (in magenta and cyan) are well separated both from the origin and from each other. Fig. 2 plots these data in the $\eta$ plane, and shows that the magenta cluster corresponds to  type I nematicity while the blue cluster is of type II. This is borne out visually in the conductance images of panels (d)-(f), in which the two major clusters form spatially separated regions. The data show phase separation between type I, typified by the the circled site of panel (e), and type II, typified by the circled site of panel (d). 

  \begin{figure*}[h]
    \centering
    \includegraphics[trim= 0 50 0 90,clip,width=0.95\textwidth]{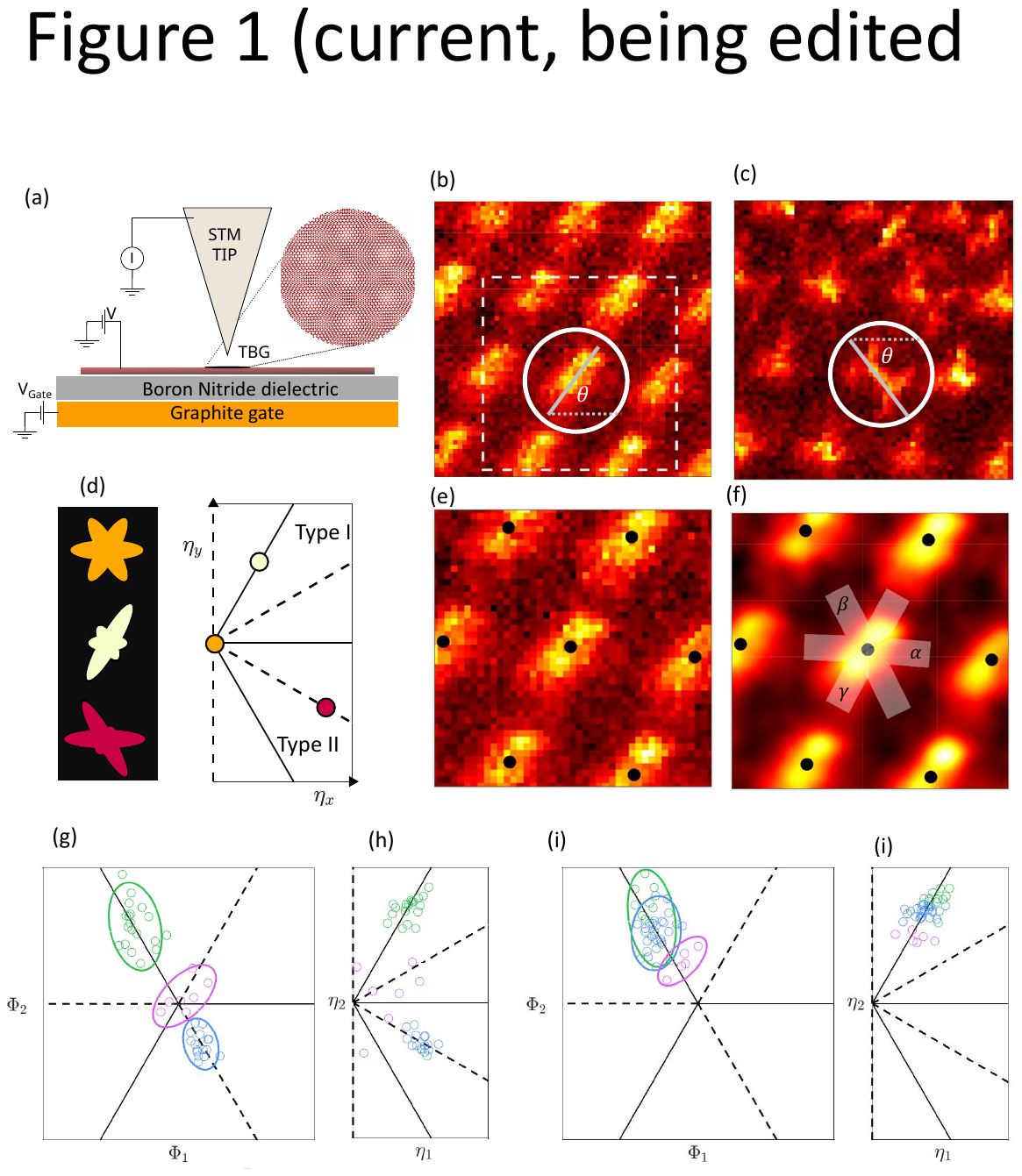}
    \caption{(a) Schematic of the experimental geometry, with illustration of the moir\'{e} pattern that entails a very large lattice constant. 
(b) STM conductance map on a hole-doped MATBG sample at a bias voltage of $+12.0$ mV, showing clear breaking of the $60^\circ$ rotation symmetry of the moir\'e lattice. The circle highlights a site with the rod-shaped conductance distribution typifying type I nematicity, with director $\eta$ (illustrated by the gray bar) at an angle $\theta \approx 60^\circ$. Data are taken from \cite{Choi2019NatPhys}, 
(c) Conductance map on a charge-neutral sample at bias $-0.8$ mV, showing a disordered pattern of rotational symmetry breaking. The circle highlights a site with the X-shaped  conductance distribution typifying Type II nematicity, with $\eta$ pointing at $\theta\approx -30^\circ$. 
(d) Illustration of types I and II nematicity. Referring to the shapes on the black background to the left: the orange shape at the top has all bonds of equal intensity, and is not nematic; the yellow shape below has one bond stronger than the others, constituting type I nematicity; the red shape at the bottom has one bond weaker than the others, constituting type II nematicity. The corresponding values of the director $\eta $ are plotted to the right. (e) Zoom of the region enclosed by the dashed rectangle in 1b. The black circles show the AA sites.(f) The region of 1e following a bilinear interpolation and a gaussian blur. In addition to the AA sites, patches along $0,\ 120$, and $240$ degrees are overlaid. The conductance in these three patches is averaged  to form the bond variables $\alpha,\beta,\gamma$. (g) GMM clustering of a synthetic dataset displaying the coexistence of type I and type II nematicity.  The data points are colored according to the cluster to which they are assigned, and the corresponding distributions are represented by $2\sigma$ ellipses; the data are represented in the $\eta$ plane in (h). (i,j) Analogous plots for a dataset exhibiting long range order of type I nematicity.}
    \label{Fig1}
\end{figure*}

  The coexistence of the two types, which are not related by any symmetry, is unexpected on theoretical grounds. Note that, because of the near equal population of data points in the two main clusters, a simple average of the data set shown would not yield a result along a high-symmetry axis, and would therefore miss this mesoscale spatial variation of well-developed nematicity.

   Near zero bias (Fig. 2g,h), the clustering assigns nearly all of the data points to a single broad cluster (blue). The corresponding values of  $\eta$ (Fig. 2h) are also very broadly distributed, as would be expected in a disorded nematic phase with spatial fluctuations of large amplitude but extremely small correlation length. At small negative voltages there are again two significant clusters, in blue and green, but both of type II character, with typical director near $-30^\circ$, apparently modeling amplitude variation of a single type. This is consistent with the predominance of x-shaped motifs (similar to the illustration in the bottom left of Fig. 1d) in the conductance images of (n)-(p), such as the site circled in (p). The dramatic bias voltage dependence of both the effective number of clusters and their underlying distributions indicates strong correlation effects among low energy excitations.
   
Under hole doping, shown in Fig. 3, the sample exhibits dramatically different behavior. In contrast to the results above, the clusters are generally contiguous, covering a range of variation of either the amplitude of the nematic order parameter or its angle. As the bias is lowered, weak type I nematicity gives way to strong, and eventually to a continuous distribution of angles with no well defined type. For large positive bias (rightmost strip in panel (a), with $\bf \Phi$ data plotted in panel (b)), the GMM fit finds three clusters that all have values of $\bf\Phi$ near the high-symmetry axis at $120^\circ$. Equivalently, as shown in panel (c), $\eta$ points near $60^\circ$ for each cluster, indicative of the nearly homogenous type I behavior obvious to the eye from the rod-like motifs in the conductance images of (d)-(f). Data for slightly smaller voltages, plotted in panels (g) and (h), exhibit similar behavior but with larger amplitude, with three clusters which, while clearly separated, all lie along the same high-symmetry line. This is consistent with a picture of long range order with spatial variation of the order parameter amplitude, as is evident in the varying intensity of the rods of (i)-(k). 
  
At small positive bias voltage, panels (l),(m), the nematicity is of a completely different character, with the directions of $\bf\Phi$ and $\eta$ much more broadly distributed, but always far from $60^\circ$. Here, there are only two clusters of significant population, both meaningfully separated from the origin, and spread between the high-symmetry directions of  $0^\circ$ and $-30^\circ$. The associated distributions overlap, and the nematic director varies continuously in angle, as can be seen in the conductance distributions in (n)-(p). Remarkably, however, this nematicity (which is of comparable strength to that exhibited at higher bias voltages), has angle completely different angle than the order parameter of the apparent long range order at larger bias. As such, we find evidence of long range order that manifests itself in a strongly bias-dependent fashion.

  The variety of phenomenology encountered in these results is quite striking considering that they are all taken within the exact same same field of view (at multiple dopings and bias voltages). Among the bias voltage windows considered, some suggest strong nematic fluctuations without any order, others suggest well-established order of type I, others of type II, and still others the coexistence of both types. This coexistence is unexpected from a theoretical perspective--the two types are not related by symmetry, and therefore in general are not degenerate. An external symmetry breaking field such as strain, which would have some fixed set of principal axes, and would be expected to definitively prefer one type over the other. This fact, as well as the broader profusion of nematic behavior, may indicate that nematicity (whether long-range ordered or not) is an intrinsic correlation effect within this region of the sample.

  To summarize, we have developed a new set of tools for the analysis of STM data on moir\'e materials that can help overcome the challenges posed by the large size of the unit cell, and have applied these tools to analyze nematic order in MATBG. The first tool uses coarse graining of conductance images to define a local nematic order parameter at each site that accords with the rotational symmetry breaking readily observable by eye. This order parameter transforms under the correct two dimensional representation of the point group, and therefore it can be used both to make contact with theoretical analysis and to capture the previously unexplored physics of the two qualitatively distinct types of nematicity. The second tool consists of the application of gaussian mixture model clustering to data sets aggregated from order parameter values at multiple sites and bias voltages. This aggregation mitigates the problem of the small number of moir\'e unit cells within the field of view, and the clustering can reveal the coexistence of multiple qualitative types of nematicity where present. Applying these tools to data from hole-doped and charge-neutral MATBG, we find indications of both types of nematic order as well as coexistence thereof, suggesting strong intrinsic nematic tendencies in this material.

   We expect both steps of our method to have broader applicability. First, the local order parameter we define can be applied to any image generated by STM (or other imaging techniques), and by virtue of its symmetry can describe any ``universal" nematic physics contained therein. Further, the coarse graining procedures we have developed can easily be applied to other forms of symmetry breaking, or in other materials\cite{Li2021}. Second, our aggregation and clustering procedure will be generally helpful when data are limited by experimental conditions, and constitutes an objective method for discovery of a range of nematic (or other symmetry breaking) behavior.

 \begin{figure*}[b]
    \centering
    \includegraphics[trim= 0 0 0 90,clip,width=1.0\textwidth]{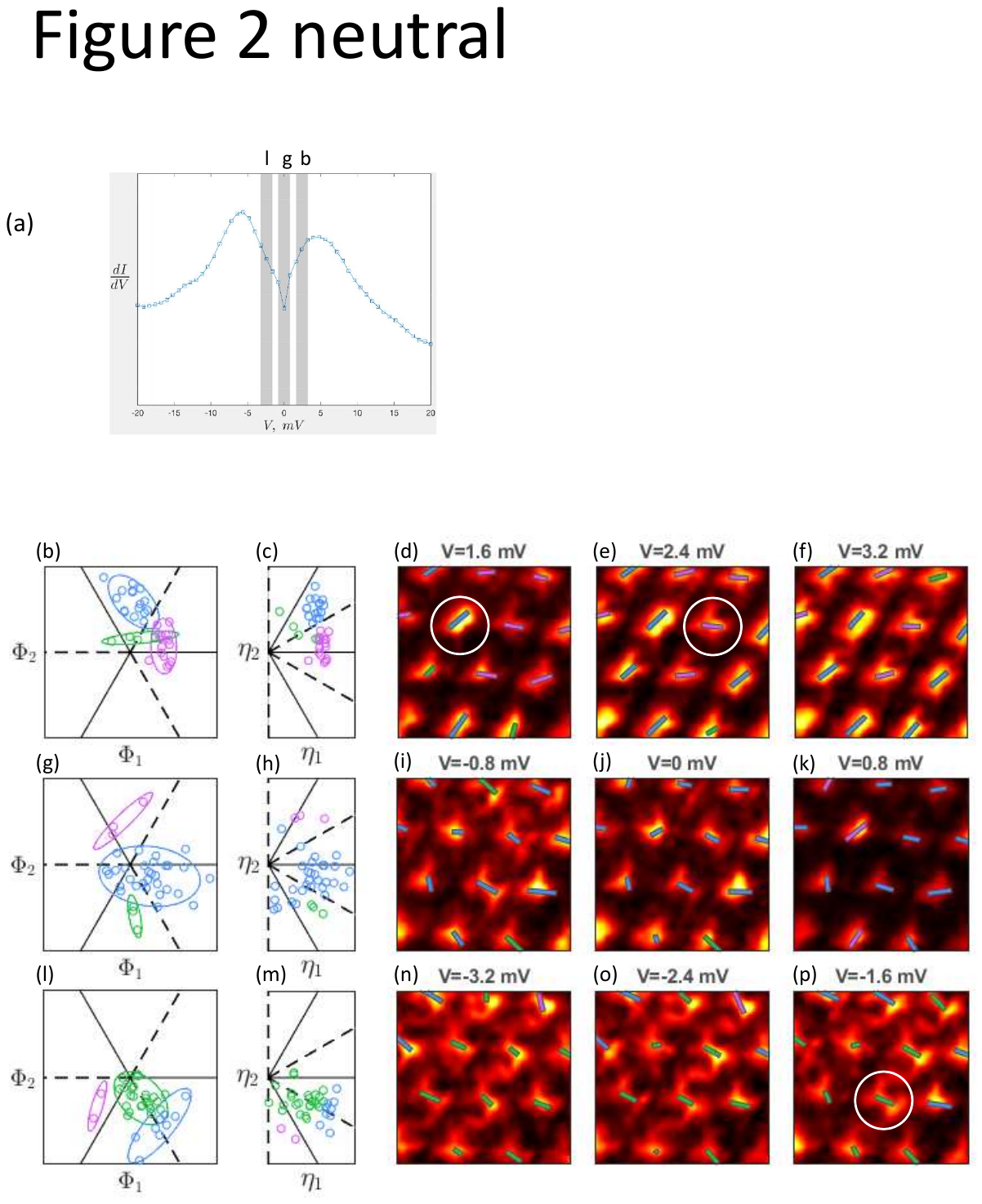}
    \caption{(a) Bias voltage dependence of the sample-averaged conductance for the charge-neutral sample of 1c. The shaded regions represent the voltage windows for the rows that follow. (b, g, l) Data for each of the selected voltage ranges plotted in the $\bf\Phi$ plane, with cluster assignments indicated by color and the fitted gaussian distributions represented as $2\sigma$ ellipses. The data are shown in the $\eta$ plane in (c,h,m). The voltage sets are, in mV: top row [-3.2,-2.4,-1.6]; middle row [-0.8,0.0,0.8]; bottom row [1.6,2.4,3.2].
(d)-(f); (i)-(k); (n)-(p): Conductance maps for each of the voltage ranges above, with values of $\eta$ shown with rods at each site and cluster assignments indicated by color.
}
    \label{fig:Fig3}
\end{figure*}

  \begin{figure*}[b]
    \centering
     \includegraphics[trim= 0 0 0 90,clip,width=1.0\textwidth]{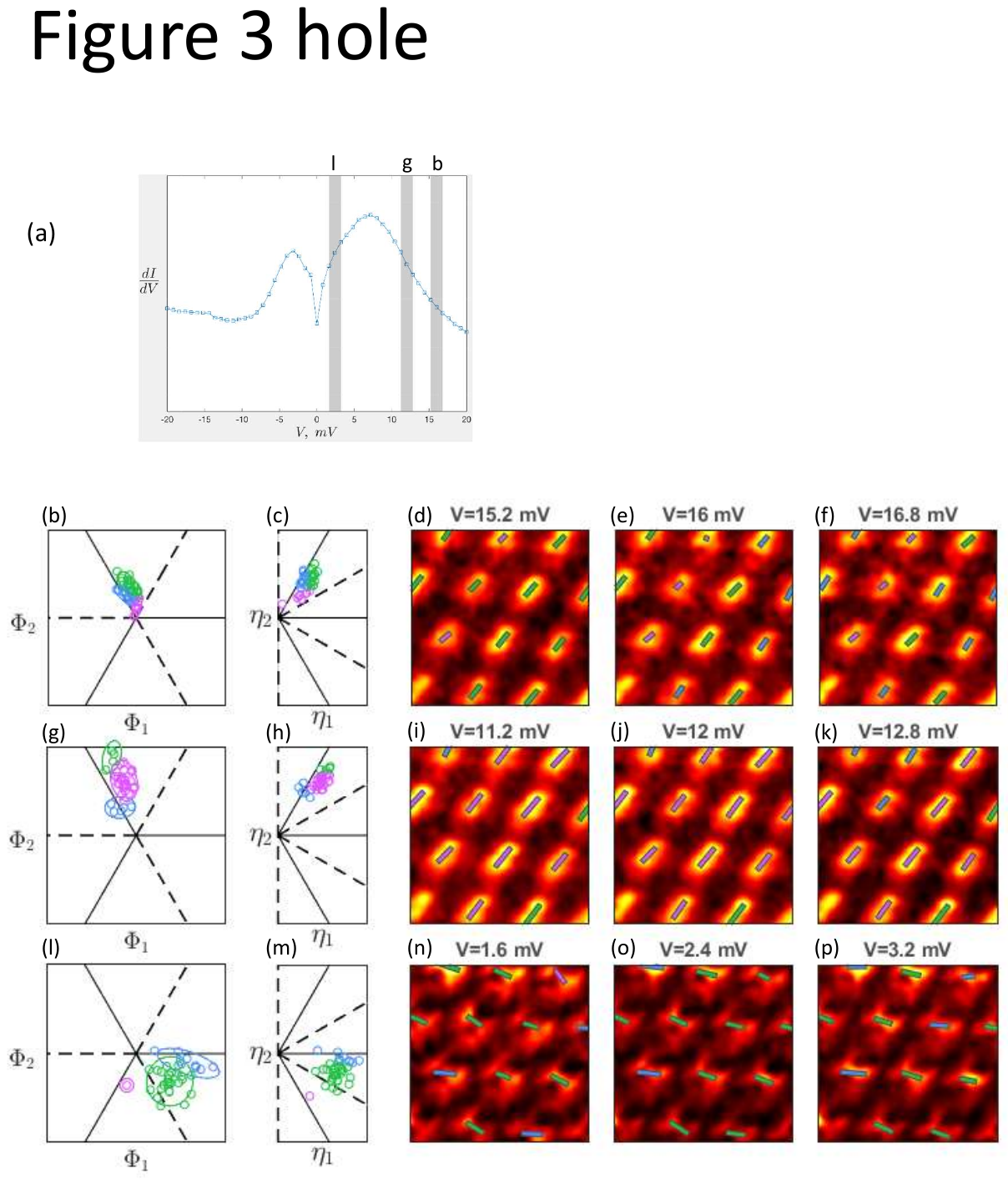}
    \caption{(a) Bias voltage dependence of the sample-averaged conductance for the hole-doped sample of 1b. The shaded regions represent the voltage windows for the rows that follow. (b, g, l) Data for each of the selected voltage ranges plotted in the $\bf\Phi$ plane, with cluster assignments indicated by color and the fitted gaussian distributions represented as $2\sigma$ ellipses. The data are shown in the $\eta$ plane in (c,h,m).The voltage sets are, in mV: top row [15.2,16.0,16.8]; middle row [11.2,12.0,12.8]; and bottom row [1.6,2.4,3.2]. (d)-(f); (i)-(k); (n)-(p): Conductance maps for each of the voltage ranges above, with values of $\eta$ shown with rods at each site and cluster assignments indicated by color.
}

\label{fig:Fig2}
\end{figure*}

{\bf Acknowledgements.}
SL,PI, AGW, and E-AK acknowledge NSF, Institutes for Data-Intensive Research in Science and Engineering – Frameworks (OAC-19347141934714). SL is supported by the U.S. Department of Energy, Office of Science, National Quantum Information Science Research Centers, Quantum Systems Accelerator (QSA). S.N-P. acknowledges support from the NSF (grant DMR-2005129) and the Sloan Foundation (grant FG-2020-13716).

\clearpage
\bibliographystyle{apsrev4-2-titles}
\bibliography{refs}

\end{document}


\title{Supplementary material for ``Unsupervised learning of two-component nematicity from STM data on magic angle bilayer graphene"}
\author{William Taranto}
\affiliation{Laboratory of Atomic and Solid State Physics, Cornell University, 142 Sciences Drive, Ithaca NY 14853-2501, USA}
\author{Samuel Lederer}
\affiliation{Department of Physics, University of California, Berkeley, USA}

\author{Youngjoon Choi}

\affiliation{T. J. Watson Laboratory of Applied Physics, California Institute of Technology, Pasadena, CA, USA}
\affiliation{Institute for Quantum Information and Matter, California Institute of Technology, Pasadena, CA, USA}
\affiliation{Department of Physics, California Institute of Technology, Pasadena, CA, USA}

\author{Pavel Izmailov}
\affiliation{Courant Institute of Mathematical Sciences, New York University, NY, USA}
\author{Andrew Gordon Wilson}
\affiliation{Courant Institute of Mathematical Sciences, New York University, NY, USA}

\author{Stevan Nadj-Perge}
\affiliation{T. J. Watson Laboratory of Applied Physics, California Institute of Technology, Pasadena, CA, USA}
\affiliation{Institute for Quantum Information and Matter, California Institute of Technology, Pasadena, CA, USA}

\author{Eun-Ah Kim}
\affiliation{Laboratory of Atomic and Solid State Physics, Cornell University, 142 Sciences Drive, Ithaca NY 14853-2501, USA}

\maketitle
\section{Transformation of $\bf\Phi$ under $D_6$}

Here we illustrate how the order parameter $\bf\Phi$ defined in the text transforms under the $E_2$ representation of the $D_6$ point group. In the main text we defined bond variables $\alpha,\ \beta,\ \gamma$ as averages over conductance regions oriented at $0,\ 2\pi/3,\ 4\pi/3$ from horizontal, with the nematic order parameter $\bf\Phi$ defined as

\begin{align}
\Phi_1=&\alpha-\frac{1}{2}\big[\beta+\gamma\big], \\
\Phi_2=&\frac{\sqrt{3}}{2}\big[\gamma-\beta\big]
\end{align}
Each element of the symmetry group $D_6$ acts as a permutation of these variables, so we will illustrate how $\bf\Phi$ transforms under a few of these. Consider the cyclic permutation $\alpha\to\gamma,\ \beta\to\alpha,\ \gamma\to\beta$, corresponding to a spatial rotation by $2\pi/3$. $\bf\Phi$ transforms as
\begin{align}
\Phi_1\to\tilde{\Phi}_1=&\beta-\frac{1}{2}\big[\gamma +\alpha\big]\\
=&-\frac{1}{2}\left(\alpha- \frac{1}{2}\big[\beta+\gamma\big] \right)-\frac{\sqrt{3}}{2}\left(\frac{\sqrt{3}}{2}\big[\gamma-\beta\big]\right)\\
=&-\frac{1}{2}\Phi_1-\frac{\sqrt{3}}{2}\Phi_2\\
\Phi_2\to\tilde{\Phi}_2=&\frac{\sqrt{3}}{2}\big[\alpha-\gamma\big]\\
= &\frac{\sqrt{3}}{2}\left(\alpha- \frac{1}{2}\big[\beta+\gamma\big] \right)-\frac{1}{2} \left(\frac{\sqrt{3}}{2}\big[\gamma-\beta\big]\right)\\
= &\frac{\sqrt{3}}{2}\Phi_1-\frac{1}{2} \Phi_2
\end{align}
or more compactly
\begin{align}
{\bf\Phi}\to\tilde{\bf\Phi}=\begin{pmatrix}-\frac{1}{2} & -\frac{\sqrt{3}}{2}\\\frac{\sqrt{3}}{2}&-\frac{1}{2} \end{pmatrix}\bf\Phi,
\end{align}
so that $\bf\Phi$ is rotated by $4\pi/3$. This is twice that of the underlying spatial rotation; it is the director $\eta$ that transforms as a (headless) vector under rotations. Similarly, under the other cyclic permutation, corresponding to a spatial rotation by $4\pi/3$, $\bf\Phi$ undergoes an $8\pi/3$ rotation.
The other class of permutations leave one bond variable invariant and switch the others, corresponding to spatial reflection symmetries. For instance, we can take  $\alpha\to\alpha,\ \beta\to\gamma,\ \gamma\to\beta$, corresponding to a reflection about the x or y-axis. For this permutation $\Phi_1\to\Phi_1$, $\Phi_2\to-\Phi_2$, so that $\bf\Phi$ also undergoes a reflection.

\section{Data processing}
\begin{figure}
         \includegraphics[width=0.3\linewidth]{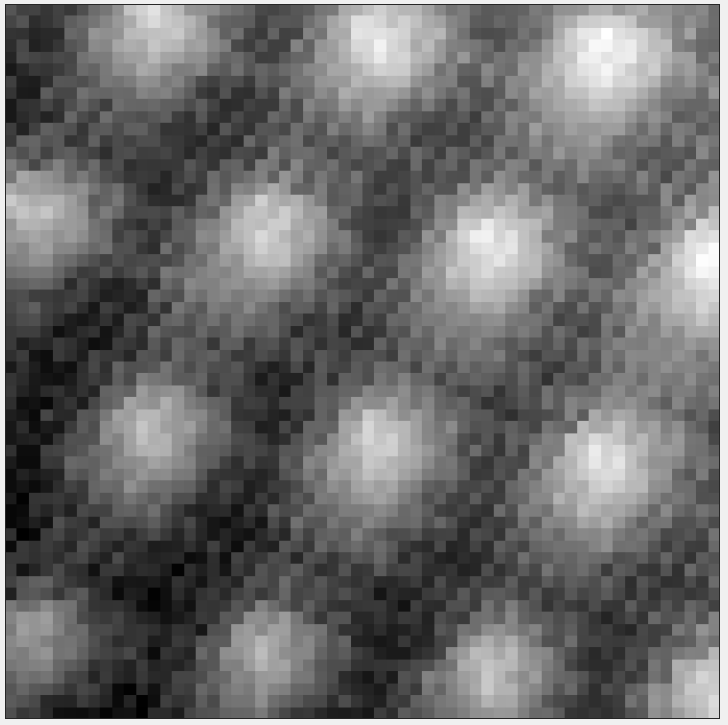}
         \caption{Raw topography data for the MATBG sample investigated in this study.}
	\label{fig:topographyRaw}
    \end{figure}

\begin{figure}
        \includegraphics[width=0.3\linewidth]{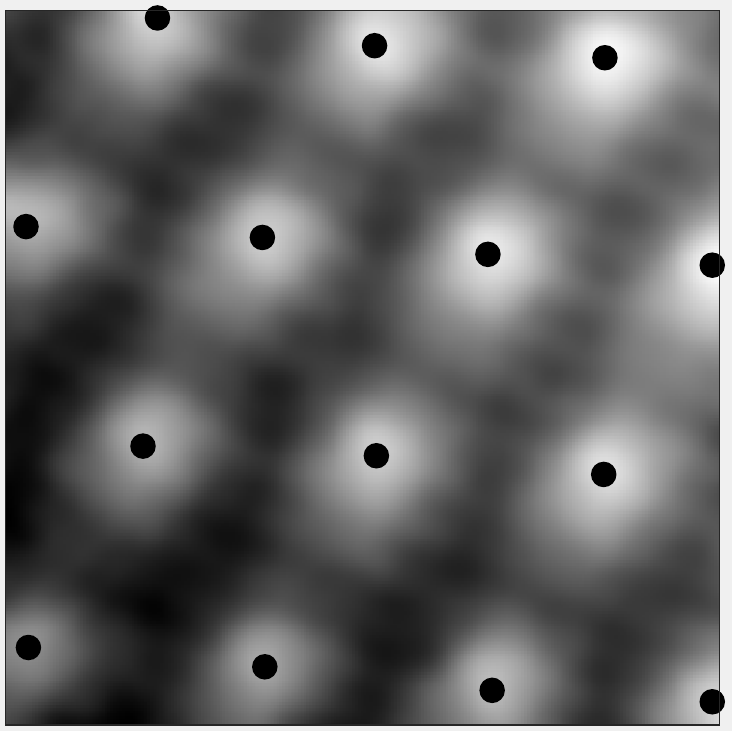}
        \caption{Topography smoothed by the method in the text, with bilinear interpolation and Gaussian blur.}
        \label{fig:topographyProcessed}
    \end{figure}

We begin with STM tunneling conductance and topography data over a field of view containing 12 moir\'e unit cells. The conductance images have high conductance regions near the local maxima of the topography, which are the AA sites of the moir\'e lattice. However, identifying the AA sites as the literal local maxima of the image may be incorrect due to outlier pixels and noise, as you can see in Fig. \ref{fig:topographyRaw}. Therefore we smooth the data as follows: first we applying a linear interpolation of the image on a four times finer grid; then apply a Gaussian filter (blur) filter with standard deviation of one pixel (of the original image). Identifying the  AA sites as the literal local maxima of the resulting image yields results in clear accord with the actual shapes of the regions of large surface height, as can be seen in Fig. \ref{fig:topographyProcessed}.

\begin{figure}
    \includegraphics[width=0.3\linewidth]{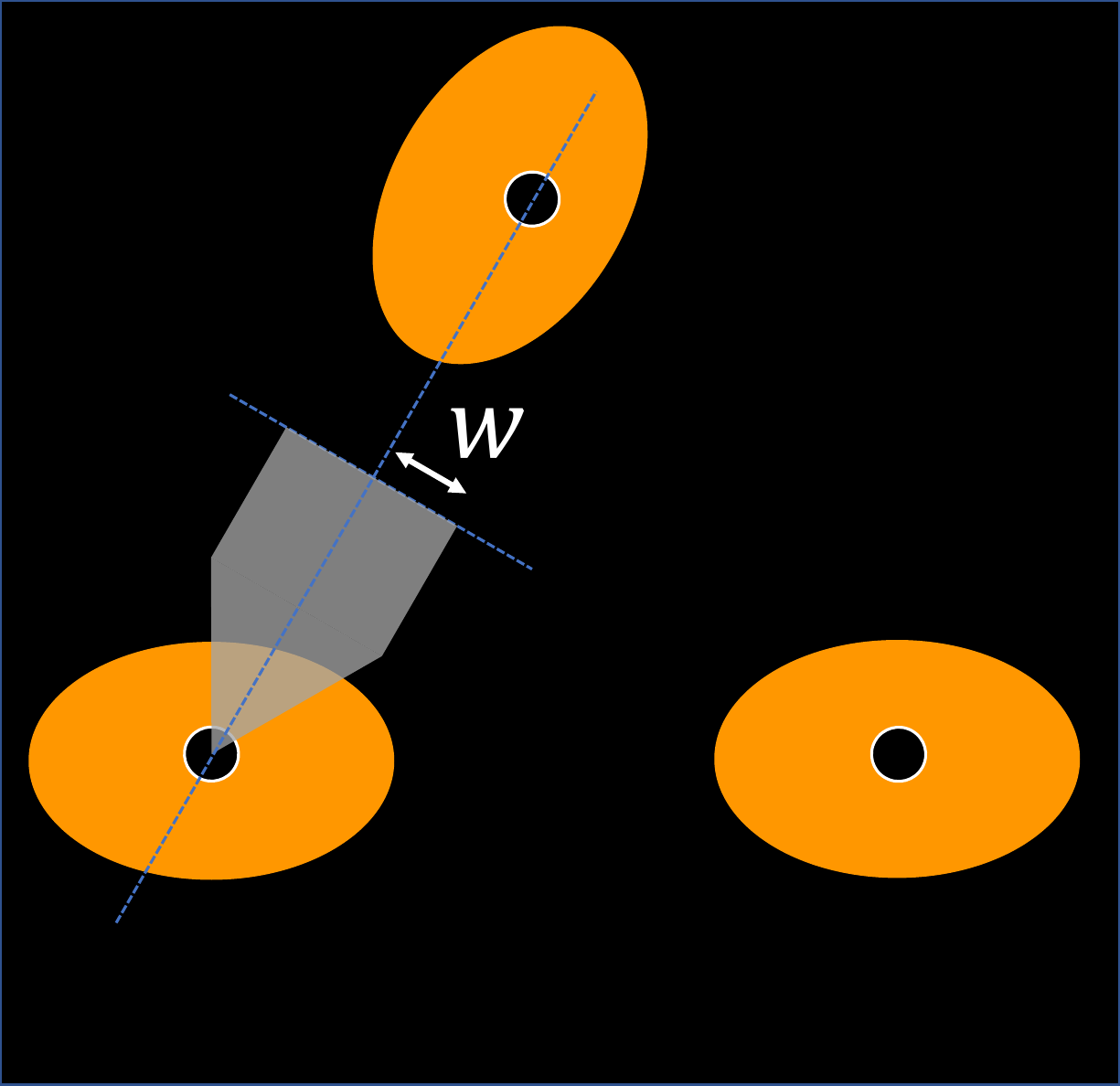}
    \caption{Illustration of the regions over which the conductance is averaged to form the bond variables used in our analysis.}
    \label{fig:propeller}
    \end{figure}

Next, we define six -shaped regions surrounding each site, in which each propeller extends towards a nearest neighbor.  The regions open at an angle of $60^\circ$, extend to the midpoint of the bond, and have a perpendicular extent $w$ on each side, as shown in Fig.\ref{fig:propeller}, where we use $w=2$ pixels in our work. To form a bond variable we average the smoothed conductance over two opposite propeller regions: $0^\circ$ and $180^\circ$ for $\alpha$, $120^\circ$ and $300^\circ$ for $\beta$,  $240^\circ$ and $60^\circ$ for $\gamma$.
%

\section{Landau theory}
The distinction between types I and II, and the phenomenological conditions under which each is preferred, can be understood from a simple Landau theory. For this purpose, it is convenient to represent the order parameter as a complex number 
\begin{align}
{\bf \Phi}=|{\bf \Phi}|
\begin{pmatrix}&\cos[2\theta]\\&\sin[2\theta] \end{pmatrix}\Longrightarrow \phi=|{\bf \Phi}|\exp[2i\theta]\label{eq:complex}
\end{align}
With this definition, we write the free energy to the lowest physical order as ${\cal F}={\cal F}_0+{\cal F}_n$, with\footnote{Note that this low-order free energy is phenomenological only. The presence of a cubic term makes the transition generically first order. Since the order parameter is not necessarily small at the transition, the expansion in $\phi$ is not actually permitted.}
\begin{align}
{\cal F}_0=& u|\phi|^2+g|\phi|^4,\\
{\cal F}_n=& \frac{\gamma}2\left(\phi^3+[\phi^*]^3\right)
\end{align}

The term ${\cal F}_0$ is the usual lowest-order Landau free energy for a single-component order parameter (such as Ising magnetism or s-wave superconductivity), for which terms odd in $\phi$ are ruled out by symmetry. However, because the nematic order has two components, there is no symmetry prohibiting the cubic term ${\cal F}_n$. If we now substitute Eq. \ref{eq:complex} into the full free energy, we obtain

\begin{align}
    {\cal F}=&u|{\bf\Phi}|^2+g|{\bf\Phi}|^4+\frac{\gamma}{2}|{\bf\Phi}|^3\left(\exp[6i\theta]+\exp(-6i\theta)\right)\\
    =&u|{\bf\Phi}|^2+g|{\bf\Phi}|^4+\gamma|{\bf\Phi}|^3\cos[6\theta]
\end{align}
The dependence on $\theta$ is entirely contained in the final term proportional to $\gamma$. If $\gamma<0$, then the system will choose $6\theta$ to be an integer multiple of $2\pi$, so that  $\theta=-\pi/3,\ 0,\ \text{or}\ \pi/3$ (or equivalents). This is type I nematicity, with the director oriented at $-60^\circ,\ 0^\circ\ \text{or}\ 60^\circ$ from the horizontal axis. Alternatively, if $\gamma>0$, the system prefers type II nematicity, with director at $-30^\circ,\ 30^\circ\ \text{or}\ 90^\circ$ from the horizontal axis. 

Within low order Landau theory, our discussion above has been completely general, relying only on the $C_3$ symmetry of the Hamiltonian. However, the appropriate coefficients in this Landau theory, including that of the crucial cubic term, depend on microscopic details. While such coefficients can be computed within a microscopic model, even qualitative results will be strongly model dependent. This makes it all the more important to quantify nematicity in experiment in such a way as to make contact with the theoretical analysis above.

\bibliographystyle{apsrev4-2}
\bibliography{refs}